\documentclass[12pt]{article}
\usepackage{epsf}
\def\reference{\parskip 0pt\par\noindent\hangindent 0.5 truecm}

\newcommand{\gesim}{\,\raisebox{-0.4ex}{$\stackrel{>}{\scriptstyle\sim}$}\,}
\newcommand{\oiii}{\hbox{[O\,{\sc iii}]}}
\newcommand{\arcmin}{\hbox{$^\prime$}}
\newcommand{\arcsec}{\hbox{$^{\prime\prime}$}}
\newcommand{\eg}{e.g.\ }
\newcommand{\ie}{i.e.\ }
\begin{document}
%
%
\title{Near-infrared micro-variability of radio-loud quasars.}
%


\author{Matthew Whiting $^{1}$ \and
 Alicia Oshlack $^{1}$ \and
 Rachel Webster $^{1}$ 
} 

\date{}
\maketitle

{\center 
$^1$ Astrophysics group, School of Physics, University of Melbourne, 
Victoria 3010, Australia\\
mwhiting,aoshlack,rwebster@physics.unimelb.edu.au\\[3mm] }

%
\begin{abstract}
We observed three AGN from the Parkes Half-Jansky Flat-spectrum Sample
at near-infrared (NIR) wavelengths to search for micro-variability. In
one source, the blue quasar PKS 2243$-$123, good evidence for NIR
micro-variability was found. In the other two sources, PKS 2240$-$260
and PKS 2233$-$148, both BL Lacertae objects, no such evidence of
variability was detected. We discuss the implications of these
observations for the various mechanisms that have been proposed for
micro-variability.
\end{abstract}



%
%

\section{INTRODUCTION}

One of the distinctive features of blazar AGN is rapid flux
variability, which is seen in every region of the electro-magnetic
spectrum (Wagner \& Witzel 1995). This variability can be on
time-scales as short as hours to minutes -- in this case, it is termed
``micro-variability''. In the optical, these variations can be of the
order of $\sim 0.1$ mag per night (see the review by Miller \& Noble
1996). While some radio-quiet quasars have been shown to exhibit
micro-variability (Jang \& Miller 1997), it is seen most commonly in
radio-loud (mostly blazar-type) AGN. The duty cycle\footnote{The
fraction of time that an object in a given class is variable. This
figure is determined observationally by taking the ratio of the time
for which objects of a given class are variable to the total observing
time for objects in that class.} for radio-loud AGN is $\sim 68\%$
(Romero et al.\ 1999) and the figure may be even higher ($\sim 80\%$)
for radio-selected BL Lacs (Heidt \& Wagner 1996).

For radio-quiet AGN, the situation is less clear. Romero et al.\
(1999) found a low duty cycle of $\sim 7\%$. Gopal-Krishna et al.\
(2000), in their extensive study of micro-variability of radio-quiet
QSOs, found that about 31\% of the objects showed evidence for
variability. They suggest that radio-quiet QSOs are less likely to
exhibit micro-variability in a given time-span than radio-loud AGN,
and that the character of the variations in the two classes tend to be
different. However, de Diego et al.\ (1998), in a comparative study of
radio-loud and radio-quiet quasars found that micro-variations may be
as common in radio-quiet quasars as radio-loud quasars. While the
micro-variations in radio-loud objects are thought to be due to shocks
within the jet, the existence of similar variability in radio-quiet
objects indicates that the accretion disk may also be important in
producing the variations.

We are conducting an on-going study of the optical and near-infrared
emission from objects from the Parkes Half-Jansky Flat-spectrum Sample
(PHFS, Drinkwater et al.\ 1997), with the data and modelling already
presented in Francis, Whiting \& Webster (2000, hereafter FWW) and
Whiting, Webster \& Francis (2001, hereafter WWF) respectively. Based
on the above estimates for micro-variability duty-cycles, it is
reasonable to expect to see micro-variability in quasars and BL Lacs
from the PHFS. This paper contains an investigation into the existence
of micro-variability in three such sources. We choose three objects
from the PHFS to search for micro-variability at near-infrared
wavelengths -- a wavelength region that has rarely been used for such
a study. With the above figures, if three objects are observed there
is a probability of $\gesim 97\%$ of seeing micro-variability in at
least one source. We also choose objects with different optical/NIR
spectra, so that we have both jet-dominated and accretion
disk-dominated sources. This will enable a comparison of the two ideas
for the production of micro-variability.

\section{OBSERVATIONS}
On the night of the 3rd of August, 1999, three AGN from the PHFS were
observed multiple times in three NIR filters over the course of $\sim
7.5$ hours, in an attempt to detect short time-scale
micro-variability. The observations were made in the near-infrared
with the CASPIR $256\times 256$ InSb array camera (McGregor et
al.\ 1994) on the Australian National University's 2.3m telescope, at
Siding Spring Observatory, NSW. The images were taken using the fast
camera with 0.5\arcsec/pixel, and thus the entire detector spans a
little over 2 arcmin.

\begin{figure}
\begin{center}
\epsfbox{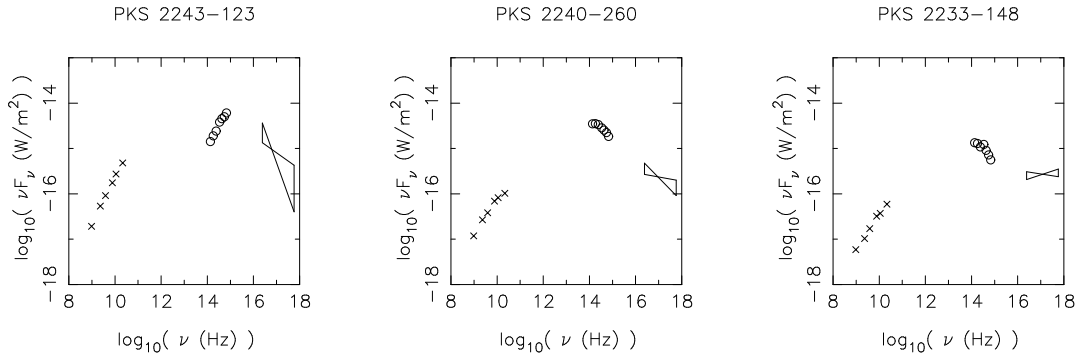}
\caption{\footnotesize Plots of the multi-wavelength SED for each of
the sources observed. The radio data (crosses) come from Kovalev et
al.\ (1999), the optical/NIR photometry (circles) is from Francis et
al.\ (2000), and the X-ray observation is from Siebert et al.\
(1998).}
\label{fig-sed}
\end{center}
\end{figure}

The three objects chosen were the following: 
\begin{itemize}
\item {\bf PKS 2243$-$123:} A quasar, $z=0.63$, that was described as
optically variable by Shen et al.\ (1998), although they did not give
a reference for this classification (it possibly comes from the
description as variable in Hewitt \& Burbidge (1993)). It has a blue
optical/NIR spectral energy distribution (SED) from FWW, with
$B-K=2.15$. WWF modelled the SEDs of many PHFS sources by fitting a
combination of a blue power law plus a synchrotron model with a
turn-over. This source was best fit with a power law of the form
$\lambda^{-1.92}$, and showed no evidence for synchrotron emission in
its optical/NIR SED. The optical spectrum of the source shows strong
broad Balmer lines and \oiii\ lines of small equivalent width
(Tadhunter et al.\ 1993).
\item {\bf PKS 2240$-$260:} A BL Lac object at $z=0.774$. It has a
redder optical/NIR SED ($B-K=4.43$ from FWW, and is fitted by WWF with
a strong synchrotron component). High optical polarisation has been
observed in this source ($15.1\%$, Wills et al.\ (1992)), and only
weak, if any, emission lines have been seen. This object was seen to
exhibit optical intra-day variability by Heidt \& Wagner (1996).
\item {\bf PKS 2233$-$148:} A candidate for a BL Lac (according to
Padovani \& Giommi (1995)), at $z>0.609$. (The redshift of $z=0.325$
quoted in Johnston et al.\ (1995) and on NED\footnote{The NASA/IPAC
Extragalactic Database, http://nedwww.ipac.caltech.edu/} appears to be
due to a mis-reading of a table in Schmidt \& Green (1983).) This has
a similar optical/NIR SED to PKS 2240$-$260, also dominated by
synchrotron emission.
\end{itemize}

\begin{figure}[t]
\begin{center}
\epsfbox{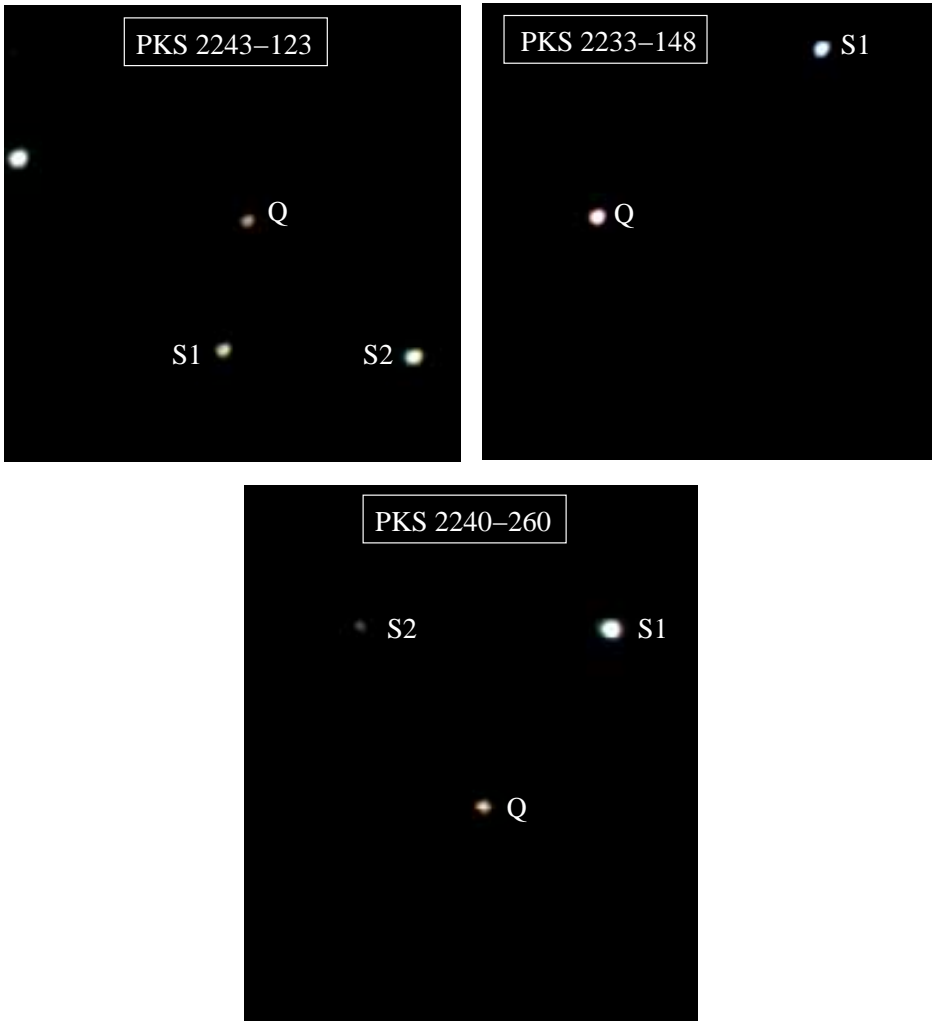}
\caption{\footnotesize Three-colour composite NIR images of the
sources used for the micro-variability analysis. The source in
question is indicated by a ``Q'', while the comparison stars
(discussed in Section~\ref{sec-var}) are indicated by ``S1'' and
``S2''. In all images, north is up and east is to the left.}
\label{fig-idv-sources}
\end{center}
\end{figure}

The multi-wavelength (radio -- X-ray) SED for each source is shown in
Fig.~\ref{fig-sed}, while three-colour images of each of the sources
are shown in Fig.~\ref{fig-idv-sources}, indicating the location of
the source, as well as nearby stars that were used in the analysis
(see Section~\ref{sec-var} for details).

These sources were chosen primarily on the basis of their position on
the sky (to enable constant observations for the second half of a
night, and to minimise the time required for the telescope to move
between each source), and also to examine the potential for
micro-variability in both BL Lacs (PKS 2240$-$260 and PKS 2233$-$148)
and blue radio-loud quasars (PKS 2243$-$123).

These objects were observed in three filters: $J$, $H$ and $K_n$. Each
source was observed eight times in each filter. Each of the images is
comprised of $2\times 2$ dithered 60 second images, each made up of
twelve averaged 5 sec exposures in $K_n$, six averaged 10 sec
exposures in $H$, and two averaged 30 sec exposures in $J$. The flat
field was created from the difference of dome exposures with the lamps
on and off -- this removes any telescope emission, improving greatly
the photometric accuracy. The sky emission was removed by using a
median of the four dithered images of the same band taken at the same
time. The four images were then aligned and combined using the median
(to remove any residual errors). The photometry was done with the {\it
apphot} package in {\sc iraf}, using circular apertures with the sky
background level determined by the median flux in an annulus around
the source. Several different sizes of aperture were used to make sure
the underlying host galaxy was not affecting the observed
variability. No significant difference was seen for the different
aperture sizes -- the results presented here use an aperture of
5\arcsec. The seeing varied from $\sim 1 - 1.2\arcsec$ over the course
of the night. The errors in the photometry shown on the light-curves
are purely the random errors from the photometry calculations.

\section{MEASURING THE VARIABILITY}
\label{sec-var}

\begin{figure}
\begin{center}
\epsfbox{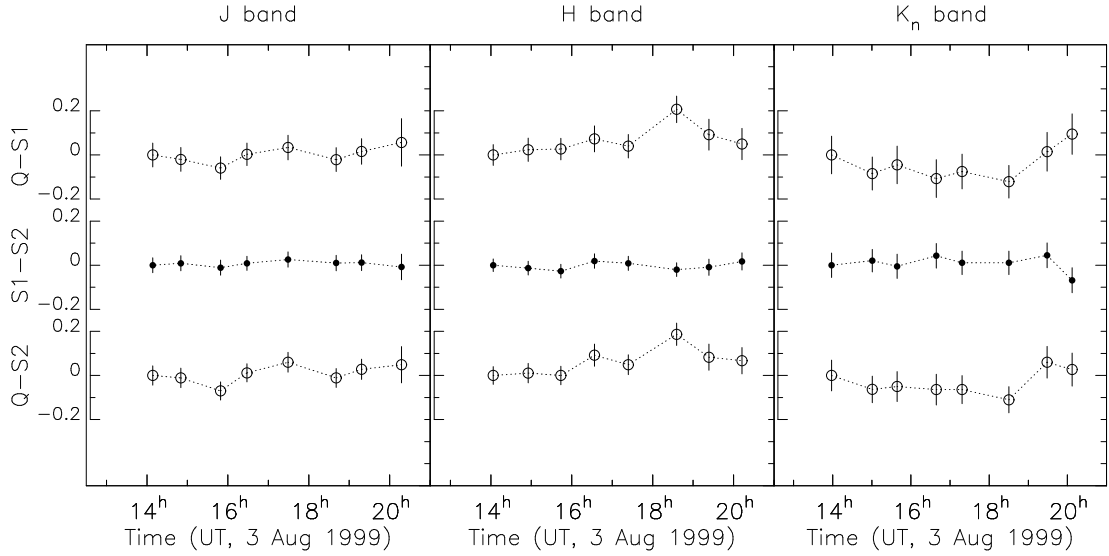}
\caption{\footnotesize Differential light-curves for PKS
2243$-$123. Shown are the quasar-star light-curves for each comparison
star (Q$-$S1 and Q$-$S2), as well as the star-star comparison
light-curve (S1$-$S2). The error bars are just the photometric
errors. The vertical scales show tick marks every 0.1 magnitudes, with
the zero point for each curve set at the first observed point. The
true values of these initial points are given in
Table~\ref{tab-idv-offset}.}
\label{fig-idv1}
\end{center}
\end{figure}

\begin{figure}
\begin{center}
\epsfbox{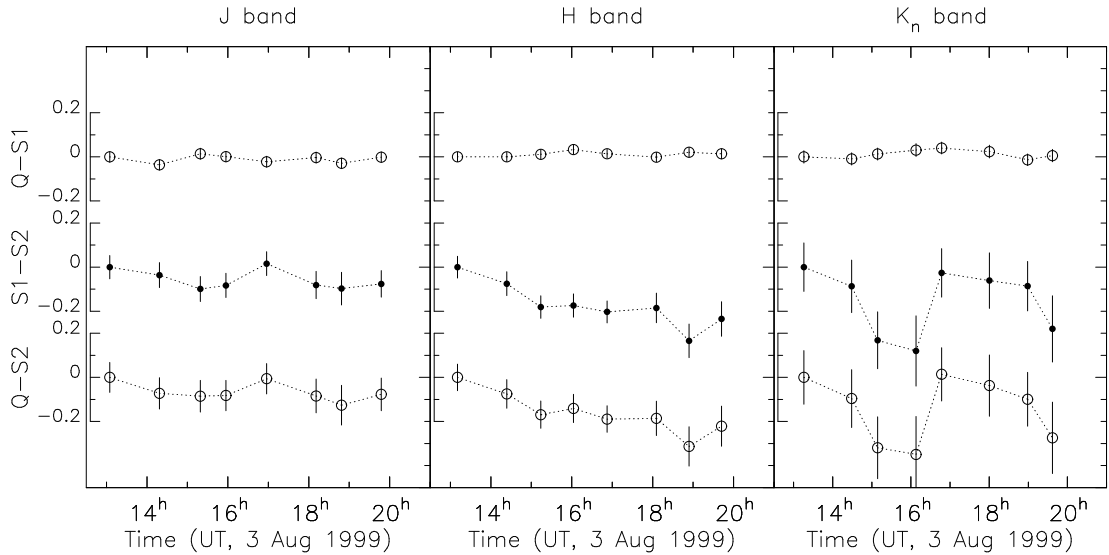}
\caption{\footnotesize Differential light-curves for PKS
2240$-$260. Details are as for Fig~\ref{fig-idv1}.}
\label{fig-idv2}
\end{center}
\end{figure}

\begin{figure}
\begin{center}
\epsfbox{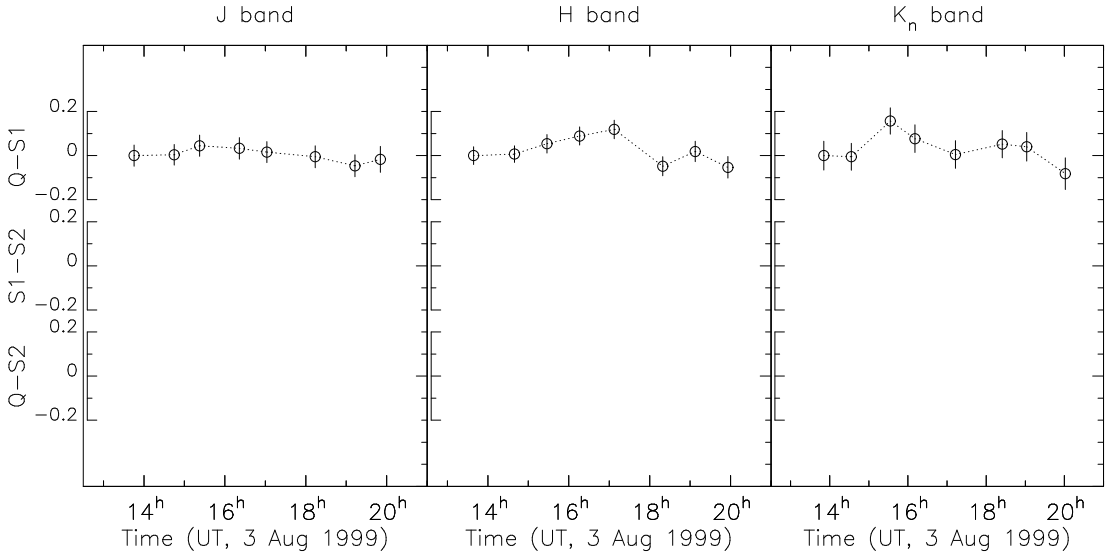}
\caption{\footnotesize Differential light-curves for PKS
2233$-$148. Details are as for Fig~\ref{fig-idv1}.}
\label{fig-idv3}
\end{center}
\end{figure}

In order to construct light-curves for each of the sources, the
differential magnitude of the quasar is measured relative to
comparison stars in the same field. This differential magnitude is
given by $\Delta m = m_q - m_c$, where $m_q$ and $m_c$ are the
magnitudes of the quasar and the comparison star respectively. This
differential magnitude is then compared to the comparison differential
magnitude $\Delta m^\prime = m_c - m_{c^\prime}$, where $m_{c^\prime}$
is the magnitude of a second comparison star in the same field. This
second differential magnitude acts as a check -- a change in flux in
the quasar will change $\Delta m$ but not $\Delta m^\prime$, while a
change that occurs in both will not be intrinsic to the quasar. The
light-curves are presented in
Figs.~\ref{fig-idv1}--\ref{fig-idv3}. For clarity, they are offset
from each other and plotted with their own vertical scale, relative to
the initial point. The zero points for these scales are given in
Table~\ref{tab-idv-offset}.

\begin{table}
\begin{center}
\begin{tabular}{ccccc} \hline
Source		&Band	&Q$-$S1	&Q$-$S2	&S1$-$S2\\ 
\hline
PKS 2243$-$123	&$J$	&0.65	&1.86	&1.21\\
PKS 2243$-$123	&$H$	&0.79	&1.73	&0.93\\
PKS 2243$-$123	&$K_n$	&0.51	&1.32	&0.81
\vspace{1mm}\\ 
PKS 2240$-$260	&$J$	&2.61	&$-1.44$	&$-4.05$\\
PKS 2240$-$260	&$H$	&2.16	&$-1.41$	&$-3.58$\\
PKS 2240$-$260	&$K_n$	&1.49	&$-2.12$	&$-3.61$
\vspace{1mm}\\
PKS 2233$-$148	&$J$	&0.37	&--	&--\\
PKS 2233$-$148	&$H$	&0.07	&--	&--\\
PKS 2233$-$148	&$K_n$	&$-0.56$	&--	&--\\
\hline
\end{tabular}
\caption{\footnotesize Offsets for the micro-variability light-curves
presented in Figs.~\ref{fig-idv1}--\ref{fig-idv3}. To obtain the
correct differential magnitude for each light-curve, add the
corresponding number from the table to the values in the figure.}
\label{tab-idv-offset}
\end{center}
\end{table}

It must be noted here that PKS 2233$-$148 did not have a second
comparison star in the small ($\sim 2\arcmin$) field of view of
CASPIR. This means that the full variability analysis cannot be done
for this source. However, a light-curve for $\Delta m$ is presented
(Fig.~\ref{fig-idv3}), using the one available star.

To determine whether a given source is variable, we use the 99\%
criterion from Jang \& Miller (1997) and Romero et al.\ (1999). We
define the variability confidence level $C=\sigma_T/\sigma$, where
$\sigma_T$ and $\sigma$ are the standard deviations of the target
(quasar--star) and comparison (star--star) light-curves. The
variability criterion requires that, for a source to be variable,
$C>2.576$. Note that these standard deviations merely come from the
point-to-point fluctuations in  the light curve, and bear no
relationship to the error bars displayed in
Figs~\ref{fig-idv1}-\ref{fig-idv3}. 

If a source is found to be variable, we also define (in a manner
similar to that of Romero et al.\ (1999)) the variability amplitude of
each light-curve by \[ A = \sqrt{(D_{\rm{max}} - D_{\rm{min}})^2 -
2\sigma^2}, \] where $D_{\rm{min}}$ and $D_{\rm{max}}$ are the minimum
and maximum of the differential light-curve respectively. Unlike the
amplitude define in Romero et al.\ (1999), this amplitude is measured
in magnitudes, and thus avoids the dependence that a relative
amplitude (\ie as a percentage) has on the normalisation of the
light-curve. Table~\ref{tab-idv} shows the values for $\sigma_T$ for
each source with respect to each comparison star, as well as the
confidence level $C$ and, when the source is variable, the amplitude
$A$ of that light-curve.

\begin{table}
\begin{center}
\begin{tabular}{ccccccc} \hline
Source	&Band	&$\sigma$  	&Comp.	&$\sigma_T$	&$C$ 	&$A$ \\ 
Name	&	&(mag)	   	&Star	&(mag)	 	&	&(mag)\\ 
\hline
PKS 2243$-$123	&$J$	&0.011	&1	&0.048	&4.56	&0.15\\
		&	&	&2	&0.053	&4.99	&0.18\\
PKS 2243$-$123	&$H$	&0.017	&1	&0.063	&3.59	&0.21\\
		&	&	&2	&0.066	&3.77	&0.21\\
PKS 2243$-$123	&$K_n$	&0.052	&1	&0.100	&1.90	&(0.28)\\
		&	&	&2	&0.071	&1.37	&(0.24)\vspace{1mm}\\
PKS 2240$-$260	&$J$	&0.081	&1	&0.022	&0.27	&--\\
		&	&	&2	&0.092	&1.13	&--\\
PKS 2240$-$260	&$H$	&0.125	&1	&0.015	&0.12	&--\\
		&	&	&2	&0.121	&0.97	&--\\
PKS 2240$-$260	&$K_n$	&0.173	&1	&0.026	&0.15	&--\\
		&	&	&2	&0.168	&0.97	&--\vspace{1mm}\\
PKS 2233$-$148	&$J$	&--	&1	&0.023	&--	&--\\
PKS 2233$-$148	&$H$	&--	&1	&0.064	&--	&--\\
PKS 2233$-$148	&$K_n$	&--	&1	&0.065	&--	&--\\
\hline
\end{tabular}
\caption{\footnotesize Micro-variability results for the three
sources. The standard deviation of the comparison curves of the quasar
with respect to each of the comparison stars, in each band.  The
values of the variability confidence level $C$ are shown where there
is more than one comparison star, and the variability amplitude $A$ is
shown for the cases where $C>2.576$.}
\label{tab-idv}
\end{center}
\end{table}

\section{CONCLUSIONS}

\subsection{RESULTS}

Near-infrared micro-variability is detected in one source, PKS
2243$-$123. There is significant variability in both $J$ and $H$
bands, while at $K_n$ band, despite both the $Q-S$ curves showing
variability (with a greater $\sigma_T$ than the other two bands),
there is also significant noise present in the comparison light-curve,
making it impossible to say at 99\% confidence that the quasar itself
is varying. It is also worth noting that we see different variability
patterns in the different bands (\eg compare the $H$ and $K_n$
light-curves). A possible cause for this is very fast time-scale
flickering that affects the different bands differently (since the
observations for the various bands are not exactly
simultaneous). These observations, nonetheless, demonstrate the
additional information acquired when multi-colour data is used in
variability analysis (something that is rarely done).

The results for the other two sources are more equivocal. Firstly, the
lack of a second comparison star for PKS 2233$-$148 means no
conclusions can be drawn about the significance of its fluctuations,
although we do note that the values of $\sigma_T$ in the $H$ and $K_n$
bands are comparable to those of PKS 2243$-$123.

Secondly, the analysis of PKS 2240$-$260 is compromised by comparison
star 2. Since both the $Q-S2$ and $S1-S2$ curves show the same
variability, it is clear that the origin of this variability is with
this star, and not the BL Lac. However, even the $Q-S1$ light-curve
shows little variability, with the $\sigma_T$ values being less than
half those of PKS 2243$-$123, indicating that the source is most
likely not varying. Since this source was observed to be variable by
Heidt \& Wagner (1996), our observation indicates that we caught the
source in a non-variable period. We do note however that the data from
Heidt \& Wagner (1996) consisted of 41 observations over 6.39 days, so
they were more likely to observe variability.

We can perform a similar analysis on Star 2 (which has the designation
0600-44255475 in the USNO-A2.0 Catalogue\footnote{See
http://vizier.u-strasbg.fr/vizier/VizieR/pmm/usno2.htx}), to quantify
the degree of its variability. The results of this analysis are given
in Table~\ref{tab-compstar}, with the $Q-S1$ curve acting as the
comparison light-curve (\ie the source of the $\sigma$
value). Clearly, this object is exhibiting strong
micro-variability. What kind of source is this? We took a noisy
spectrum of this source with the ANU 2.3m DBS, and it indicates the
source is roughly a late K-type star. This is consistent with the
colours of the star in both the optical and near-infrared (taken from
the images from FWW). We do note that it is thus too red to be an RR
Lyrae star, which would have accounted for the rapidity of its
variations. We tentatively suggest it is instead a flare star.

\begin{table}
\begin{center}
\begin{tabular}{ccccccc} \hline
Source	&Band	&$\sigma$  	&Comp.	&$\sigma_T$	&$C$ 	&$A$ \\ 
Name	&	&(mag)	   	&Object	&(mag)	 	&	&(mag)\\ 
\hline
PKS 2240$-$260 S2	&$J$	&0.022	&Q	&0.092	&4.22	&0.34\\
			&	&	&S1	&0.081	&3.72	&0.30\\
PKS 2240$-$260 S2	&$H$	&0.015	&Q	&0.121	&8.21	&0.44\\
			&	&	&S1	&0.125	&8.45	&0.45\\
PKS 2240$-$260 S2	&$K_n$	&0.026	&Q	&0.168	&6.50	&0.43\\
			&	&	&S1	&0.173	&6.70	&0.44\\
\hline
\end{tabular}
\caption{\footnotesize Micro-variability results for the second
comparison star in the PKS 2240$-$260 field.}
\label{tab-compstar}
\end{center}
\end{table}

\subsection{INTERPRETATION}

The origin of optical and NIR micro-variability in blazar AGN is not
very well understood. There are two main types of models that have
been developed to explain the observations. Firstly, there are models
that use a shock-in-jet setup. This type of model has been used to
explain the radio variability seen in blazars (Marscher \& Gear
1985). To produce the optical micro-variability, a thin relativistic
shock within a jet encounters a feature, such as a bend in the jet or
an inhomogeneity in the density, which causes an enhancement of the
observed jet emission (see, for example, Gopal-Krishna \& Wiita (1992)
or Qian et al.\ (1991)).

A second suggestion for the cause of micro-variability moves away from
the jet and looks instead at the accretion disk. Perturbations or
instabilities on the surface of an accretion disk (such as
``hot-spots'' -- regions of high temperature and/or density) can
create micro-variations in the emission (see Mangalam \& Wiita (1993)
and references therein). It is likely that both models are important
for micro-variability in AGN, and the relative importance for
individual sources depends on the relative strengths of the jet and
disk emission.

The quasar PKS 2243$-$123 has a blue optical/NIR SED (the photometry
from these observations gives the same NIR slope as the photometry
presented in FWW, and is modelled in WWF by a blue power law). It thus
shows no evidence for synchrotron emission in its optical spectrum (a
fact also supported by its observed strong emission lines). Hence, it
is unlikely that fluctuations in the jet emission will cause the
observed micro-variations, unless the jet is present at a very low
level and experiences relatively large fluctuations. It seems then
that the micro-variability in this source is more likely to be
explained by the accretion disk instability model.

To test this statement, we can take the model fitting from WWF to an
extreme, so that a synchrotron component is fit along with a very blue
($f_\nu \propto \nu^{-0.05}$) power law. This generates the maximum
amount of synchrotron flux according to this model. The amount of
synchrotron as a proportion of total flux in the near-infrared ranges
from 3.4\% (at $J$ band) to 5.4\% (at $K_n$ band). If we assume that the
synchrotron component is the only component that is varying, then to
produce the observed fluctuations, the synchrotron component must vary
in flux by a factor of at least 7 (on a time-scale of about an
hour). The accretion disk instability model allows us to avoid such
extreme variability, and is consistent with observations of
micro-variability in radio-quiet quasars (\eg Gopal-Krishna et al.\
2000).

Finally, since the observations presented here were limited in scope
(due to the time available), a follow-up study would be quite
worthwhile. Such a study should involve selecting a small sample of
quasars (and BL Lacs) that has some sources with synchrotron dominated
SEDs (this part of the sample would contain the BL Lacs, based on the
modelling in WWF), and some with blue power law SEDs (in the manner of
PKS 2243$-$123). The observations would ideally cover a greater period
of time than that described in this paper, to increase the probability
of seeing variability. This would also enable a better determination
of the nature and structure of the variability.  The relative amounts
of micro-variability in the two types of sources could then be gauged
accurately, particularly in light of our detection of
micro-variability in the blue quasar. This would hopefully give an
indication of how important each model is for micro-variability in all
types of radio-loud AGN.

\section*{Acknowledgements}


We wish to thank the two anonymous referees for their helpful
comments, and the MSSSO TAC for granting us the time on the 2.3m
telescope.

\section*{References}





\reference de Diego, J.A. et al. 1998 ApJ 501, 69
\reference Drinkwater, M.J. et al. 1997, MNRAS 284, 85
\reference Francis, P.J., Whiting, M.T. and Webster, R.L. 2000, PASA
17, 1, 56 (FWW)
\reference Gopal-Krishna et al. 2000 MNRAS, 314, 815
\reference Gopal-Krishna and Wiita, P.J. 1992, A\&A 259, 1992
\reference Heidt, J. and Wagner, S.J. 1996, A\&A 305, 42
\reference Hewitt, A. and Burbidge, G. 1993, ApJSS 87, 451
\reference Jang, M. and Miller, H.R. 1997, AJ 114, 565
\reference Johnston, K.J. et al. 1995, AJ, 110, 880
\reference Kovalev, Y.Y. et al. 1999, A\&AS 139, 545
\reference Mangalam, A.V. and Wiita, P.J. 1993, ApJ 406, 420
\reference Marscher, A.P. and Gear, W.K. 1985, ApJ 298, 114
\reference McGregor, P., Hart, J., Downing, M., Hoadley, D., and
Bloxham, G. 1994, in Infrared Astronomy with Arrays: The Next
Generation, ed. McLean, I.S. (Dordrecht: Kluwer), p.299
\reference Miller, H.R. and Noble, J.C. 1996, in Blazar Continuum
Variability, ed. Miller, H.R., Webb, J.R. and Noble, J.C., ASP
Conference Series vol.110, 17
\reference Padovani, P. and Giommi, P. 1995, MNRAS 277, 1477
\reference Qian, S.J. et al. 1991, A\&A 241, 15
\reference Romero, G.E., Cellone, S.A. and Combi, J.A. 1999, A\&AS
135, 477
\reference Schmidt, M. \& Green, R.F. 1983, ApJ, 269, 352
\reference Shen, Z. et al. 1998, AJ 115, 1357
\reference Siebert J. et al. 1998, MNRAS 301, 261
\reference Tadhunter, C.N. et al. 1993, MNRAS 263, 999
\reference Wagner S.J. and Witzel, A. 1995, ARA\&A 33, 163
\reference Whiting, M.T., Webster, R.L. and Francis, P.J. 2001, MNRAS,
323, 718 (WWF)
\reference Wills, B.J. et al. 1992, ApJ 447, 139

\end{document}